\documentstyle[aps,rotate,multicol,epsf]{revtex}
\def\beginwide{
        \end{multicols} \vspace*{-0.5cm} \noindent
        \rule{3.5in}{.1mm}\rule{.1mm}{5mm} \widetext \medskip }
\def\beginwidetop{
        \end{multicols} \vspace*{-0.5cm} \noindent
        \widetext \medskip }
\def\endwide{
        \hspace*{3.35in}~\rule[-5mm]{.1mm}{5mm}\rule{3.5in}{.1mm}
        \begin{multicols}{2} \vspace*{-1.0cm} \noindent }
\def\endwidebottom{
        \begin{multicols}{2} \vspace*{-1.0cm} \noindent }
\def\bx{{\bf x}}

\draft

\begin{document}
\title{
A Critical ``Dimension'' in a Shell Model for Turbulence.
}
\author{Paolo Giuliani and  Mogens H. Jensen }
\address{Niels Bohr Institute, 
Blegdamsvej 17, DK-2100 Copenhagen {\O}, Denmark }
\author{Victor Yakhot}
\address{Institute for Advanced Studies, Einstein Drive, 
Princeton, NJ 08540, USA}
\date{\today}
\maketitle
\begin{abstract}
We investigate the GOY shell model within the scenario of a
critical dimension in fully developed turbulence. 
By changing the conserved quantities, one can continuously
vary an ``effective dimension'' between $d=2$ and $d=3$. We identify a critical
point between these two situations where the flux of energy changes sign
and the helicity flux diverges.
Close to the critical point 
the energy spectrum exhibits a turbulent scaling regime 
followed by a plateau of thermal equilibrium. We identify scaling laws 
and perform a rescaling argument to derive a relation between
the critical exponents. 
We further discuss the distribution function of the energy flux.
\end{abstract}
\begin{multicols}{1}

\smallskip
\smallskip
Many theoretical and experimental results for
fully developed turbulence have been offered 
over the last decade. A new approach
has been presented by Yakhot \cite{Yakhot} in which the method
of generating functions by Polyakov \cite{Pol} is generalized to the
Navier-Stokes equations. Applying a renormalization group
procedure \cite{ors} results in 
an estimate of a critical dimension for turbulence,
around $d_c \sim 2.5$, thus following the foot steps
of an original idea by Frisch and Fournier \cite{FF} but correcting 
the actual value
of the dimension. The physical idea behind the existence
of a critical dimension is related to
the well known fact that the energy cascade in three
dimensional turbulence is ``forward'' (in $k$-space) going from large
to small scales whereas for two dimensional turbulence it is backward,
from small to large scales. This leads to the identification of a critical
dimension between two and three at which the flux of energy changes its
sign, and the amplitude of the field turns into a peak where
there is no flux neither forward nor backward. 
In ref. \cite{Yakhot} the theory is expanded around this critical point
in terms of a ratio between two time scales.
However, it is not possible
to investigate the physical behavior in a non-integer dimension directly,
neither experimentally nor numerically. 
In this letter we therefore propose to study this type
of criticality in a shell model for turbulence \cite{mogens}.
In particular we focus on the GOY model \cite{Gledzer,YO,JPV}
which exhibits well known
conservation laws: in the 3-$d$ version energy and helicity 
are conserved; in the 2-$d$ version energy and enstrophy are conserved. It
is possible to {\it continuously} vary the effective dimension of the model by
changing the second conserved quantity from a helicity to an enstrophy
quantity. As the energy is {\it always} conserved, we can study the energy
flux directly as a function of the variation in the second conserved quantity
and we identify a critical point, where the flux changes sign.
Indeed the second conserved quantity is non-physical at this point
as expected. Nevertheless we are able numerically to extract a series
of new properties of the spectrum and the PDF around this critical point.  
A similar observation of a change of sign in the energy flux as a 
function of a parameter was already
made in a different shell model by Bell and Nelkin \cite{Mark}. In their
model the dynamics is not intermittent and the properties of the model
are thus quite different from the GOY model. 

Our starting point is the approximative approach to turbulence made by
discretizing the wave number space by exponentially separating ``shells''
\cite{mogens}.
In this respect, we apply the GOY model \cite{Gledzer,YO} 
which has been successful in
giving results for intermittency corrections in agreement with
experiments \cite{JPV} (for other results on the GOY model, see
\cite{pisarenko,kada1,bif,kada2}). 
The starting point is a set of wave numbers $k_n = k_0
2^n$ and an associated complex amplitude $u_n$ of the velocity field.
Each amplitude interacts with nearest and next-nearest neighboring shells
and the corresponding set of coupled ODE's takes the form:
\begin{eqnarray}
\label{un}
(\frac{d}{ dt}+\nu k_n^2 ) \ u_n \ & = &
 i \,k_n (a_n \,   u^*_{n+1} u^*_{n+2} \, + \, \frac{b_n}{2} 
u^*_{n-1} u^*_{n+1} \, + \, \nonumber \\
& & \frac{c_n }{4} \,   u^*_{n-1} u^*_{n-2})  \ + \ f \delta_{n,n_f},
\end{eqnarray}
with $n=1,\cdots N$,  $k_n=r^n \, k_0$ ($r=2)$,  and  boundary conditions
$ b_1=b_N=c_1=c_2=a_{N-1}=a_N=0$.
The values of the coupling constants are fixed by imposing
conserved quantities. By conserving the total energy
$\sum_n |u_n|^2$ when $f=\nu = 0$,
we obtain the constraints
$a_n+b_{n+1}+c_{n+2}=0$.
The time scale is fixed by the condition $a_n =1$
leaving free the parameter $\delta$ by defining the coupling constants
as
\begin{equation}
a_n=1 \qquad b_n=-\delta  \qquad c_n=-(1-\delta)~~~.
\end{equation}
The model also possesses a second conserved quantity of the form
\begin{equation}
Q=\sum k_n^{\alpha} |u_n|^2
\label{q}
\end{equation}
which leads to a relation between $\alpha$ and $\delta$: 
$2^{\alpha}=1/(\delta-1)$.
For $\delta<1$ this relation requires complex values of $\alpha$,
with $\Im{(\alpha)}=\pi/\ln\,2$.
In 3d turbulence helicity 
$H=\int ({\bf \nabla \times u}(\bx)) \cdot {\bf u} (\bx)  d{\bx}$
is conserved, which in terms of shell 
variables takes the form \cite{kada1}
\begin{equation}
H=\sum_n (-1)^n \, k_n \, |u_n|^2~~~~,
\end{equation}
when the values of parameters are $\delta = \frac{1}{2}$ and
$\Re{(\alpha)}=1$. In 2d turbulence 
on the other hand enstrophy 
$\Omega=\int \vert {\bf \nabla \times u}(\bx) \vert^2 d{\bx}$
is conserved and this corresponds
to the parameters $\delta = \frac{5}{4}$, $\alpha=2$
which on the shells takes the form
$\Omega=\sum k_n^2 |u_n|^2$.
Note that energy is conserved for any value of $\delta$.
This gives us the possibility to {\it continuously} vary 
the effective dimension (and thus the generalized helicity/enstrophy
(\ref{q})) by varying the parameter $\delta$
between $\delta= \frac{1}{2}$ (3d) and $\delta = \frac{5}{4}$  (2d). 
\begin{figure}[htb]
\narrowtext
\vbox{
{
        \centerline{\epsfxsize=7.0cm
        {\epsfbox{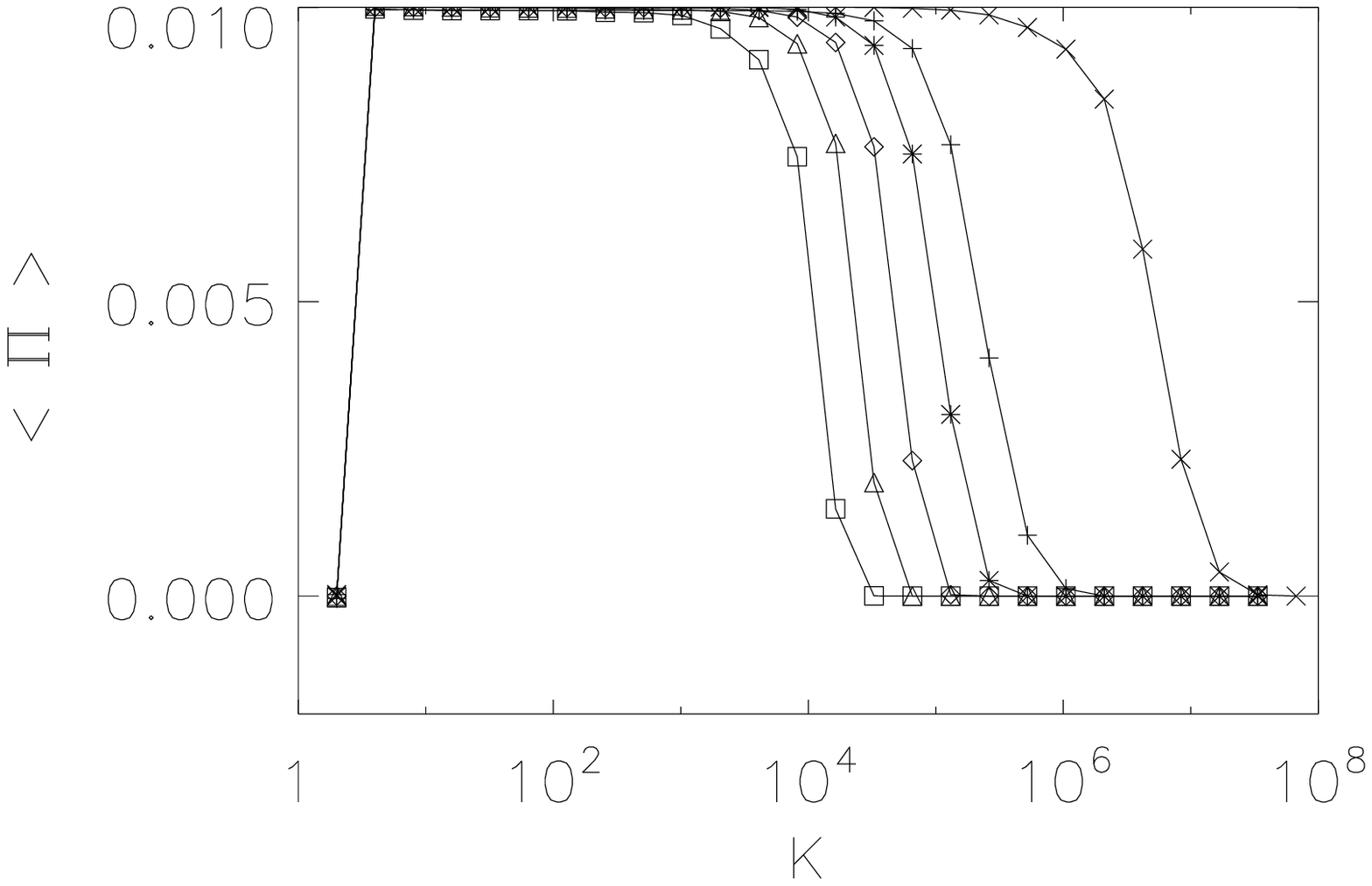}}}
}
{
        \centerline{\epsfxsize=7.0cm
        {\epsfbox{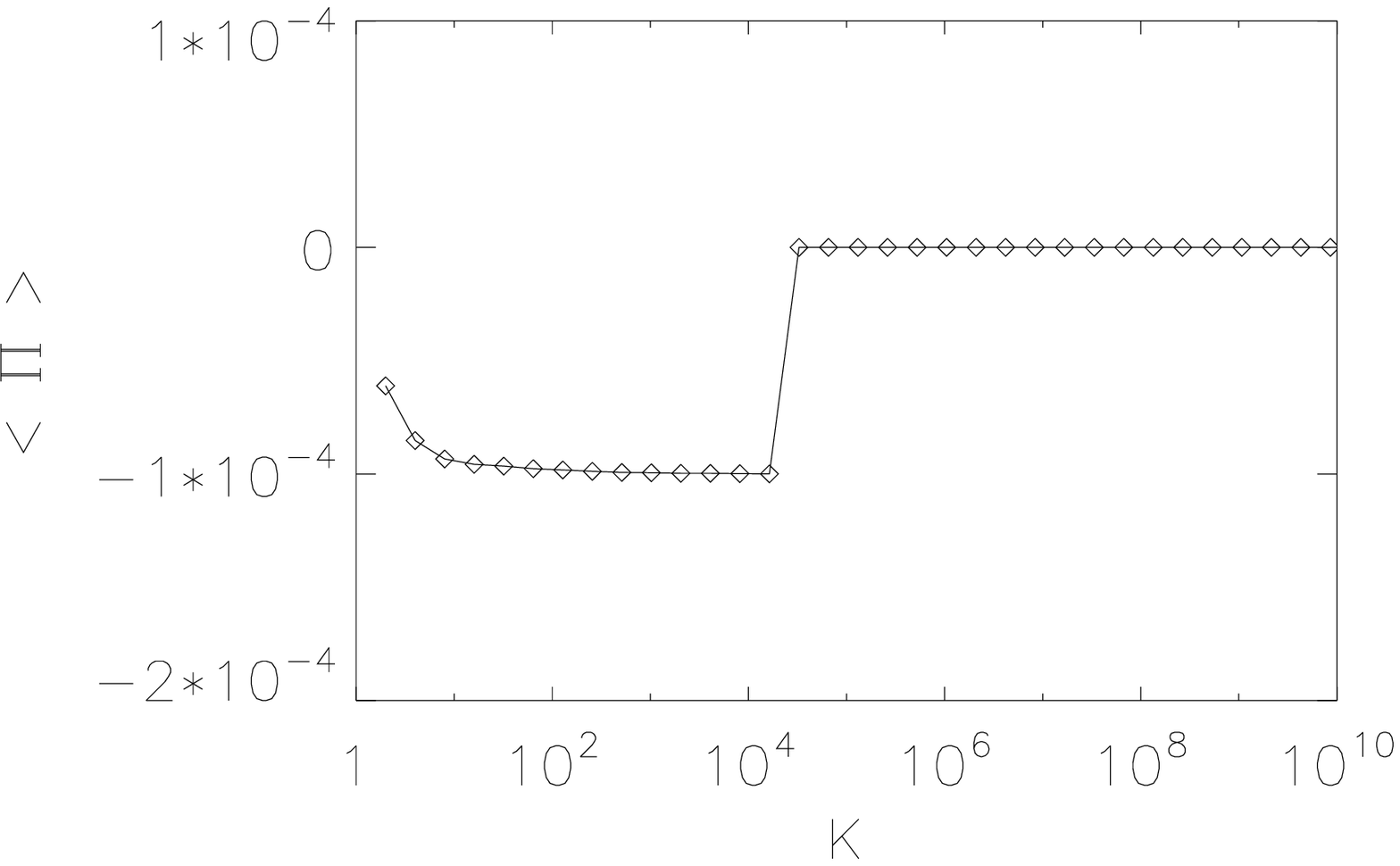}}}
}
}
\caption{
a): Average energy flux versus the wavenumber $k$ with 
$N=25, \nu = 10^{-10}, n_f = 2$ for $\delta=0.5
(\times)$, and
$\epsilon=2 \cdot 10^{-4} (+)$, $2 \cdot 10^{-5} (\ast)$, 
$\ldots$, $2 \cdot 10^{-8} (\Box)$.
Note that as $\epsilon$  decreases the inertial range shrinks. 
b):
Inverse energy flux with $N = 33, \nu = 10^{-16}$
for $\epsilon = - 0.125$. The forcing term is on
shell 15 and of the form $f_{15}=(1+i)*10^{-4}/u_{15}^{\ast}$. 
A large scale viscosity is now applied, see the text.}
\label{fluxes}
\end{figure}

The critical point is identified by looking at the energy flux
through each shell which is given by \cite{pisarenko}
\begin{eqnarray}
\Pi_n & = & 
\left\langle \, -\frac{d}{dt} \sum_{i=1}^n |u_i| ^2 \, \right\rangle  
= \nonumber \\
& & \left\langle - Im \left( k_n u_n u_{n+1} ( u_{n+2} + {(1-\delta) \over 2}
u_{n-1})) \right) \right\rangle,
\label{p_n}
\end{eqnarray} 
where only the contributions of the nonlinear terms are considered
in the time-rate-of-change of the cumulative energy.
From Eq. (\ref{p_n}) we see that the last term vanishes as 
$\delta \to 1$
causing a depletion in the energy transfer \cite{bif}. This 
is observed in the numerical simulations
shown in Fig. \ref{fluxes}a, where
the inertial range of the flux shrinks as $\epsilon \equiv 1 - \delta \to 0$.
Note, that we in Fig. 1a apply a forcing on the form
$f_{n_f} = (1+i)  \ast  10^{-2} / u_{n_f}^{\ast}$
in order to ensure a constant input of the energy (and thus a
constant flux). We reach similar conclusions 
when a constant deterministic forcing is applied. 
Moving above $\delta =1$, 
the energy flux reverses going instead from small 
to large scales, see Fig. \ref{fluxes}b. Therefore
the point $\delta = \delta_c =1$ defines a critical point where
the energy flux for finite value of energy input $f$ discontinuously jumps
from positive to negative values (the jump diminishes with 
the forcing amplitude $f$). 
According to Eq. (\ref{q}), the
generalized helicity $Q$ diverges at $\delta = 1$ (i.e.
$\alpha\to\infty$) and this could be a reason
for the inhibition of the energy transfer \cite{Lesieur}. 
Furthermore,
the rate of injected generalized helicity, given by
$\sum_n(-1)^{n}k_n^{\alpha}\Re\langle f_n u_n^{\ast}\rangle$,
diverges at this point \cite{fo}.

\begin{figure}[htb]
\narrowtext
\vbox{
{
        \centerline{\epsfxsize=7.0cm
        {\epsfbox{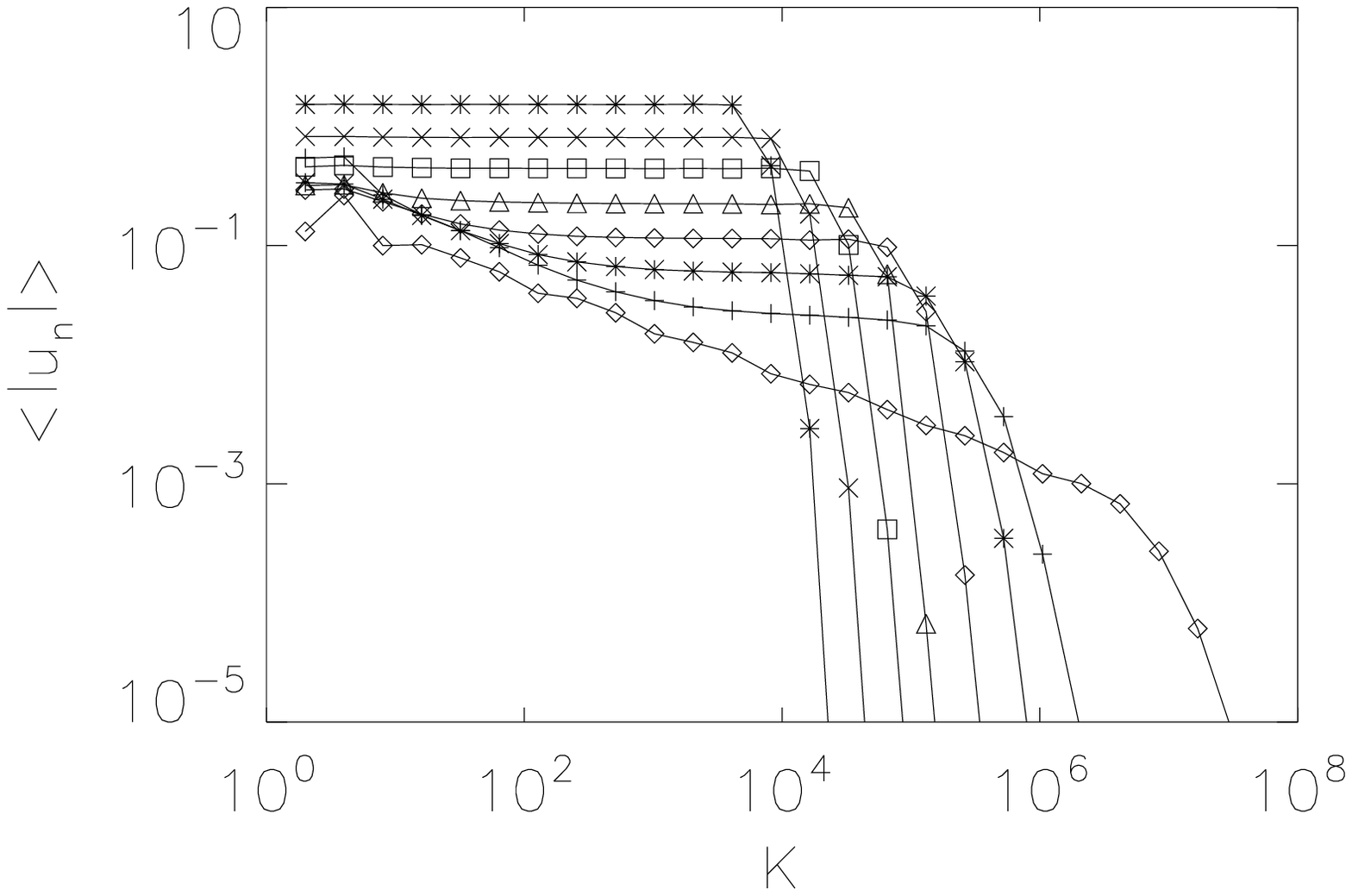}}}
}
{
        \centerline{\epsfxsize=7.0cm
        {\epsfbox{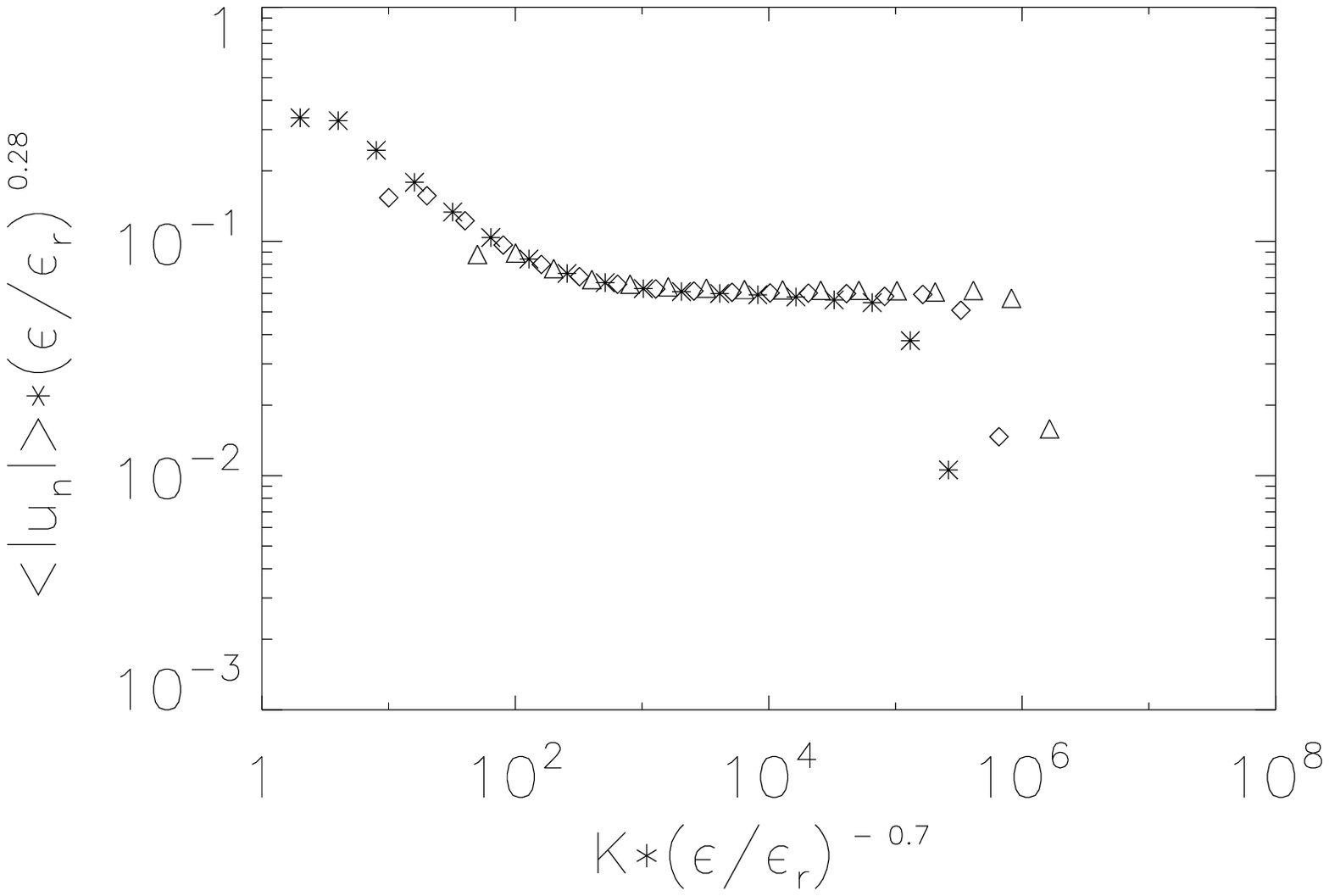}}}
}
}
\caption{
a): The spectrum $<\vert\,u\,\vert>$ versus $k$ for the 3d case
$\delta=0.5 (\diamond)$ and for 
$\epsilon= 2 \cdot 10^{-4} (+)$, $\ldots$, 
$\epsilon = 2 \cdot 10^{-10} (\ast)$. Note the flat part 
of the spectrum developing as $\epsilon \to 0$.
b): A rescaling plot of the spectra in Fig. (2)
by $ < |\,u\,| >\!\!*({\epsilon\over\epsilon_r})^{0.28}$ versus
$k\!*\!({\epsilon\over\epsilon_r})^{-0.7}$
($\epsilon\!=\!10^{-6}(\diamond)$ and $\epsilon=10^{-7}(\bigtriangleup)$),
where $\epsilon_r=2\cdot 10^{-5}(*)$. 
}
\label{fig2}
\end{figure}

Let us now turn to the spectra for $\delta<1$.
As $\delta$ is increased from 
$\delta=0.5$ and $\epsilon \to 0$, one observes
a build up of a shoulder leading to a
plateau in the spectrum at large values of $k_n$.
%
%
This is shown in Fig.\ref{fig2}, where the parameter $\epsilon$
is varied one decade for each spectrum
(note that in our spectra
we plot $< | u_n | >$ vs. $k_n$ which provides similar information
as plotting $< | u_n |^2 >$). 
We identified various scaling laws associated
with the spectra of Fig.\ref{fig2}.
First of all the dissipative cut-off, $k_d$,
moves in a systematic
fashion as a function of $\epsilon$. We find the following scaling law:
$k_d \sim \epsilon^{\alpha_d}$, with $\alpha_d \simeq 0.3$.
The plateau in the spectra (equipartition of energy among the shells)
may be interpreted as a thermal equilibrium which overcomes the
turbulent regime when the forward transfer of energy is reduced.
We found that the level of the plateau scales with $\epsilon$ as
$\langle \vert\,u_n\,\vert\rangle_{pl}\sim \epsilon^{-\alpha_{pl}}$,
where $\alpha_{pl}\simeq 0.28$, thus finally turning into a diverging
amplitude around the forcing scale.
The turbulent, cascading, part of the spectrum varies like
$\langle | u_n | \rangle \sim k^{-\alpha_s}$, $\alpha_s\simeq 0.4$,
taking into account corrections due to intermittency \cite{JPV}.
Finally, the critical wavenumber $k_c$, at which the spectrum
crosses over from turbulent behavior to thermal equilibrium,
also moves with $\epsilon$, possibly like $k_c\sim \epsilon ^{\alpha_c}$.
To determine $k_c$, we balance the contributions from the two regimes
\begin{equation}
\langle | u_n | \rangle \sim \langle \vert\,u_n\,\vert\rangle_{pl}
\Rightarrow k^{-\alpha_{s}} \sim \epsilon^{-\alpha_{pl}}~~~~,
\end{equation}
and obtain the following scaling law for $k_c$
\begin{equation}
k_c \sim \epsilon ^{\alpha_c},\alpha_c=\alpha_{pl}/\alpha_{s}\simeq 0.7~~~~,
\end{equation}
showing that the scaling exponents are not all independent \cite{note}. 
This result can be
verified by a simple rescaling of data. Let us assume that
$\langle | u_n | \rangle / \langle | u_n |\rangle_{pl}$
is a function of $k/k_c$ alone, i.e.
\begin{equation}
\frac{\langle | u_n | \rangle}{\langle | u_n | \rangle_{pl}}
\sim f({k\over k_c})~~~~,
\end{equation}
where $f(x)$ is such that $f(x)\sim x^{-\alpha_s}$, $x<<1$ and
$f(x)\sim const$, $x>>1$. Then a data collapse is obtained by plotting
$\langle | u_n | \rangle / \epsilon^{-\alpha_{pl}}$
versus
$k/\epsilon ^{\alpha_c}$. A good rescaling plot is obtained,
see Fig.\ref{fig2}b, when the estimated value $\alpha_c=0.7$ is used.
Notice that the collapse of data does not apply to the dissipative
range, since $k_d$ and $k_c$ scale differently with $\epsilon$.

\begin{figure}[htb]
\narrowtext
\vbox{
{
        \centerline{\epsfxsize=7.0cm
        {\epsfbox{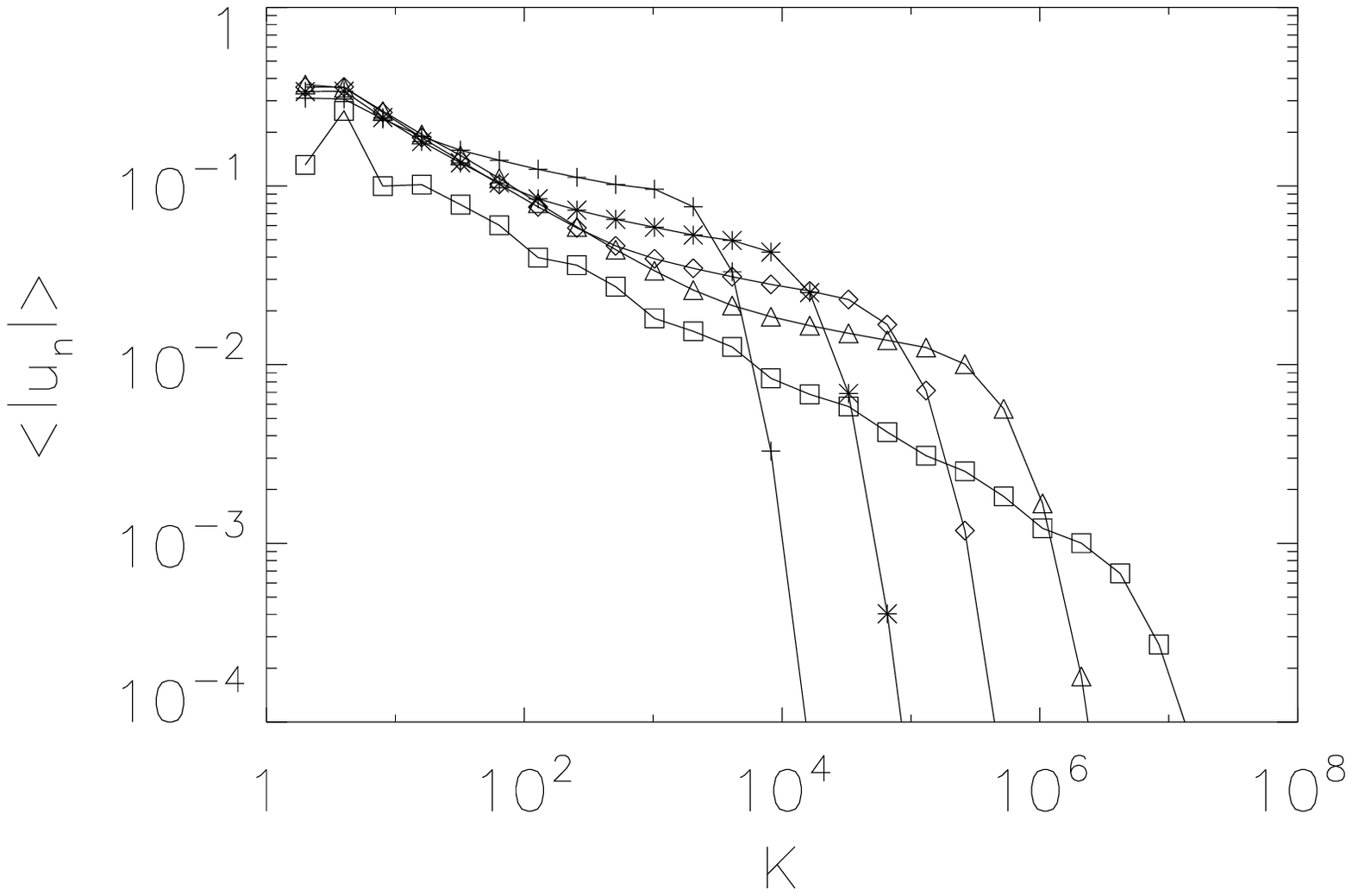}}}
}
{
        \centerline{\epsfxsize=7.0cm
        {\epsfbox{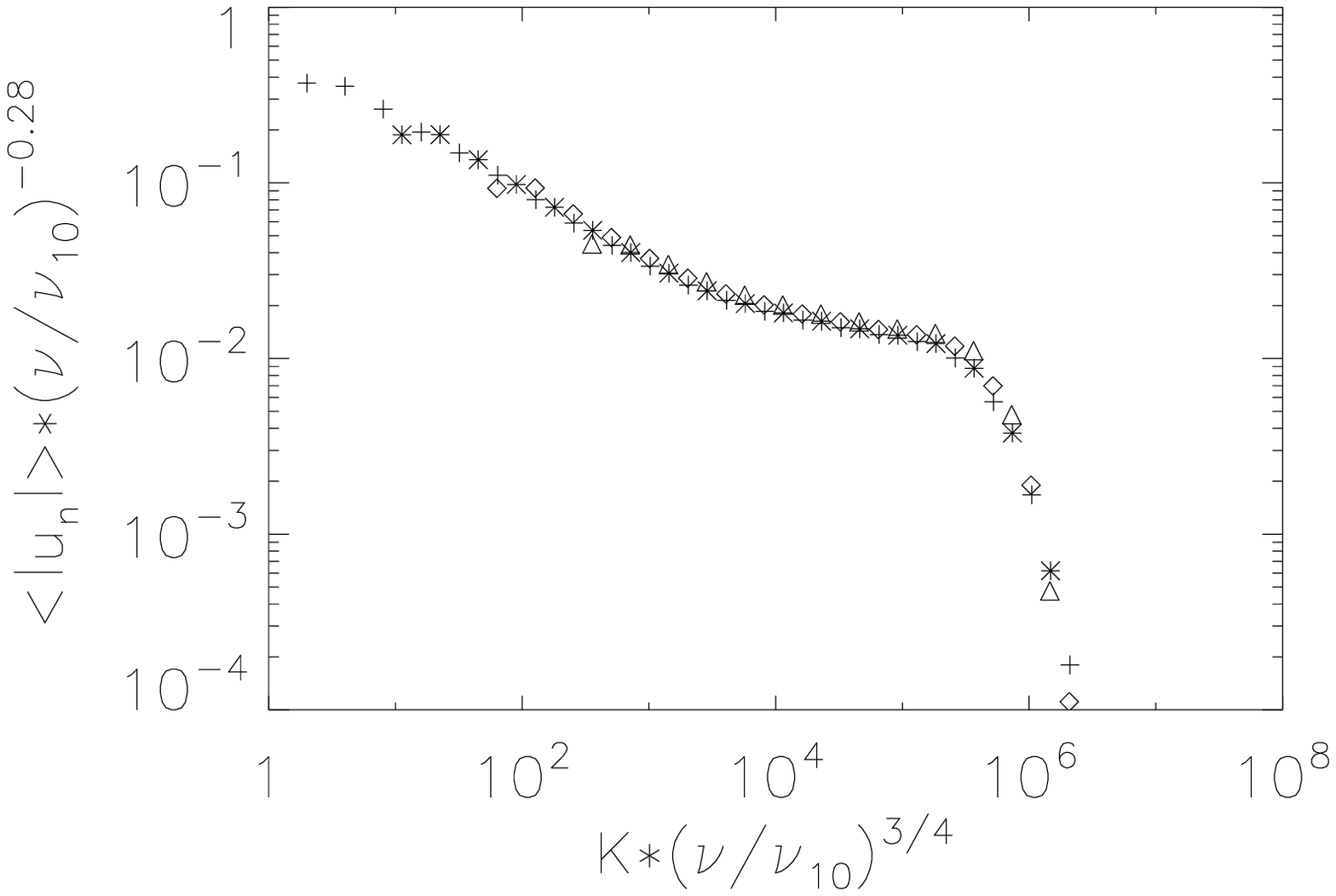}}}
}
}
\caption{
a): The spectrum $<\vert\,u\,\vert>$ versus $k$ for the standard 3d case -
$\delta=0.5, \nu = 10^{-10} (\Box) $ - and for 
$\epsilon = 2 \cdot 10^{-3}$ and $\nu =
10^{-10} (\bigtriangleup), \ldots,  10^{-7} (+)$.
b): A rescaling of the curves in a) by
$<\vert\,u\,\vert>\ast (\nu / \nu_{10})^{-0.28}$ versus
$k \ast ((\nu / \nu_{10})^{3/4}$ where $\nu_{10}= 10^{-10}$.
}
\label{n}
\end{figure}

Since there is no transfer of energy at
the critical point, the
non-linear terms will not play any role and the equations will only
include the dissipation and forcing terms. This can be made 
quantitative by the fact that the state with a peak at the forcing
scale
\begin{equation}
{\bf u} = (0,0,0,\frac{f}{\nu k_{n_f}^2 },0,0....,0)
\label{fp}
\end{equation}
is a fixed
point of the equations (\ref{un}) which at 
$\epsilon =0$ is marginally stable \cite{jul}. 
Indeed we find 
numerically at $\epsilon =0$ that by starting 
with the fixed point (\ref{fp})
the peak stays at the forcing scale and the amplitudes remain zero
above but become non-zero, although small, below the forcing scale.
On the contrary, for $\epsilon \to 0^+$, the peak is unstable
and the energy is soon redistributed to the neighbouring shells.
The existence of a sharp transition into the critical point was already
indicated by a calculation of the maximal Lyapunov exponent which
drops sharply as $\epsilon \to 0$ \cite{jul}. 
In order to study the behavior of the spectrum around the critical point in
details we have two parameters to vary, the ``dimension'' parameter, $\delta$,
and the viscosity, $\nu$. Fig. \ref{n} shows a series of spectra for 
$\epsilon = 0.002$ 
varying $\nu$. Again one observes the shoulder at large $k$. As expected, 
the shoulder moves to higher $k$ when the viscosity decreases. The wavevector
for the dissipative
cut off $k_D$ moves in the Kolmogorov fashion 
$k_d \sim \nu^{-3/4}$. This leads us again to perform a standard
finite size rescaling plot, rescaling the $k$ axis by 
$(\nu/\nu_{10})^{3/4}$ and the velocity
axis by $(\nu/\nu_{10})^{-0.28}$, see Fig. \ref{n}.

\begin{figure}[htb]
\narrowtext
\vbox{
{
        \centerline{\epsfxsize=7.0cm
        {\epsfbox{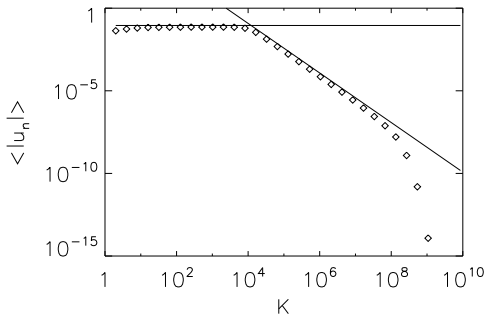}}}
}
}
\caption{
$<\vert\,u\,\vert>$ versus $k$ for $\delta=1.125$ ($\alpha=3$).
The two branches of statistical equilibrium 
$<\vert\,u\,\vert>$ $\sim const$
(energy equilibrium) and $<\vert\,u\,\vert>$ $\sim k^{-3/2}$ (generalized
enstrophy equilibrium) are clearly visible.
}
\label{2dim}
\end{figure}

Now consider the other side of the critical point, 
namely on the ``two-dimensional'' side for $\delta > 1$.
The behavior of 2d shell models has been previously investigated
in several papers \cite{Frick,Aureletal,Ditlevsen}.
A kind of ``coupled GOY model''
\cite{Aureletal}
gives an inverse flux of energy which is explained
in terms of a mean diffusive drift in a system close to statistical
equilibrium, and
shell models for 2d do not seem to give an
inverse energy cascade with the usual ``5/3'' spectrum. 
In order to extract the energy of the inverse cascade we need
to add a large scale viscosity to Eqs. (\ref{un}) of the type
$-\nu' k_n^{-2} u_n$.
In Fig. (\ref{2dim}) we show the
energy spectrum for the case $\delta=1.125, \nu = 10^{-16},
\nu' = 10^{-1}$. The two branches of statistical
equilibrium, respectively energy and generalized enstrophy 
(\ref{q}) equilibrium are clearly visible (see \cite{Ditlevsen}
for details).

\begin{figure}[htb]
\narrowtext
\vbox{
{
        \centerline{\epsfxsize=7.0cm
        {\epsfbox{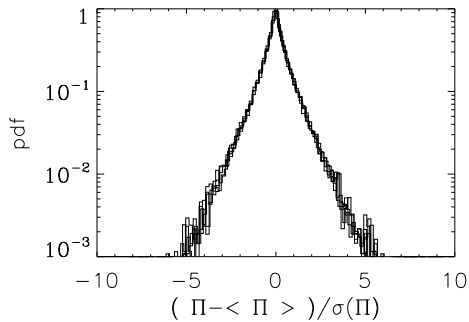}}}
}
}
\caption{
Rescaling of the PDF obtained for shells in the ``flat'' part
of the spectrum in Fig. 2 with $\nu = 10^{-10}, 
\epsilon = 3 \cdot 10^{-8}$, for the shells $n=4,6,8,10,12$. The pdf's
are in each case rescaled by the standard
deviation. 
}
\label{pdf}
\end{figure}

Let us turn our attention to the probability density functions
(PDF). It
is well known that, in fully developed turbulence, the 
PDF at the largest
scales (small $k$)
typically behaves like a Gaussian, slowly changing its form as one
moves towards the small scales (large $k$), turning into a shape
where large events play an important role giving a kind of stretched
exponential PDF. Such a non trivial behavior is related to the property
of multiscaling of the structure functions. When 
a turbulent scaling regime is detectable as in Fig. \ref{n},
we observe that the intermittency corrections appear
to persist, even close to $\epsilon = 0$. However,
in the flat part of the spectrum of Fig. \ref{fig2}, 
the probability distribution
becomes wider at larger $k$ but such that the shape is
simply rescaled onto a universal and almost symmetric
curve (rescaling by the standard deviation), see Fig. \ref{pdf}.

In this Letter we have presented results for the existence of a critical
point in the ``GOY'' shell model. Our main results can be summarized
as follows. At this critical point, which lies between three- and
two-dimensional behavior,  the energy flux changes its sign, going from 
a forward to a backward transfer. Approaching this point from
the ``three-dimensional'' side, part of the spectrum becomes flat
as an indication of a thermal equilibrium. 
The cross-over to the flat part is determined by balancing 
the turbulent energy spectrum with a spectrum in thermal equilibrium.
We identify the scaling behavior of the cross-over point and rescale
the spectra accordingly. The PDF of the thermal equilibrium shows
simple scaling invariance although the statistics is not Gaussian. 
For analytical understanding of these results, one
rewrites the ``GOY'' equations in terms of
a generating function technique \cite{Yakhot,Pol}
thus obtaining a set of coupled ordinary differential equations
\cite{PD}. One
can further map these equations onto a Fokker-Planck equation for the 
distribution of the exponentiated quantities. We will discuss
this in a forthcoming publication.

We are grateful to P. Ditlevsen, G. Eyink, U. Frisch, P. Hohenberg, P. Olesen,
A. Polyakov, I. Procaccia, B. Shraiman, K. Sreenivasan for 
interesting discussions.  We are also
indebted to the ITP, Santa Barbara and the 
program on ``Hydrodynamics Turbulence'', where this work started.

\end{multicols}

\begin{thebibliography}{10}
\bibitem[*]{*}
Electronic Address: mhjensen@nbi.dk
\bibitem{Yakhot}
V. Yakhot, Phys. Rev. E {\bf 63}, 026307  (2001).
%
\bibitem{Pol}
A.M. Polyakov, Phys. Rev. E {\bf 52}, 6183 (1995).
\bibitem{ors}
V. Yakhot and S.A. Orszag, Phys. Rev. Lett. {\bf 57}, 1722 (1986).
%
\bibitem{FF}
J.D. Fournier and U. Frisch, Phys. Rev. A {\bf 17}, 747 (1978).
%
%
%
\bibitem{mogens}
T. Bohr, M. H. Jensen, G. Paladin, and A. Vulpiani, ``Dynamical systems
approach to turbulence'', Cambridge University Press, Cambridge (1998).
%
%
\bibitem{Gledzer}
E. B. Gledzer, Sov. Phys. Dokl. {\bf 18}, 216 (1973).
%
\bibitem{YO}
M. Yamada and K. Ohkitani, J. Phys. Soc. Japan {\bf 56}, 4210(1987); Prog.
Theor. Phys. {\bf 79},1265(1988).
%
\bibitem{JPV}
M. H. Jensen, G. Paladin, and A. Vulpiani, Phys. Rev. A {\bf 43}, 798 (1991).
%
\bibitem{Mark}
T.L. Bell and M. Nelkin, Phys. Fluids {\bf 20}, 345 (1977).
\bibitem{pisarenko}
D.~Pisarenko, L.~Biferale, D.~Courvasier, U.~Frisch, and M.~Vergassola,
Phys. Fluids A {\bf65}, 2533 (1993).
%
\bibitem{kada1}
L. Kadanoff, D. Lohse, J. Wang, and R. Benzi,
Phys. Fluids {\bf 7}, 617 (1995).
\bibitem{bif}
L. Biferale, A. Lambert, R. Lima, and G. Paladin.
Physica D {\bf80}, 105 (1995).
\bibitem{kada2}
L. Kadanoff, D. Lohse, and N. Sch\"orghofer, Physica D {\bf 100}, 165 (1997).
\bibitem{Lesieur}
J.C. Andr\'e and M. Lesieur, J. Fluid Mech. {\bf 81}, 187 (1977).
\bibitem{fo}
This is a forcing dependent feature as it is possible
to adjust the forcing to keep both energy and helicity
input fixed.   
\bibitem{note}
In the spirit of ref. \cite{Yakhot}, one can do
a similar estimation for the Navier-Stokes eqs.
of the wave number at which there is a transition
from turbulent and thermal fluctuations. 
The velocity field is devided into two parts 
$u_n = v_n + V_n$
where $v_n$ relates to the statistically turbulent field and $V_n$
to the statistically thermal/equalibrium field. From \cite{Yakhot}
we know that
$v_n^2 \approx (\delta_c - \delta)^{-\frac{1}{3}} k_n^{-\frac{5}{3}}$.
Since the thermal spectrum is flat
$
V_n^2 \approx C(\delta)
$
one finds the following form of the energy spectrum
$
E(k_n) = \frac{1}{(\delta_c - \delta)^{\frac{1}{3}}}
k^{-\frac{5}{3}} + C(\delta).
$
Since the turbulent spectrum is infared divergent the main contribution
to the energy will be associated the forcing shell $n_f$. The 
equilibrium spectrum is on the other hand dominated by the
dissipation scale $k_d$ so the total energy is
$
E = \frac{1}{(\delta_c - \delta)^{\frac{1}{3}}}k_{n_f}^{\frac{2}{3}} +
C(\delta) k_d
$.
As the dissipation scale will vary like $k_d \approx 
(\delta_c - \delta)^{\frac{1}{4}}$ then 
$C(\delta) \approx (\delta_c - \delta)^{-\frac{7}{12}}$ which
leads to the following expression for the spectrum
$
E(k_n) = \frac{1}{(\delta_c - \delta)^{\frac{1}{3}}}
k^{-\frac{5}{3}} + O((\delta_c - \delta)^{-\frac{7}{12}}
,$
$
k_c \approx ((\delta_c - \delta)^{\frac{3}{20}}. 
$
For $k > k_c$ the thermal equilibrium spectrum exceeds the
turbulent spectrum.
\bibitem{jul}
J. Kockelkoren, F. Okkels and M.H. Jensen, Journ. Stat. Phys. {\bf 93},
833 (1998).
\bibitem{Frick}
P. G. Frick, and E. Aurell, Europhys. Lett. {\bf 24}, 725, (1993).
%
\bibitem{Aureletal}
E. Aurell, G. Boffetta, A. Crisanti, P. Frick, G. Paladin, and A. Vulpiani,
Phys. Rev. E {\bf 50}, 4705, (1994).
%
\bibitem{Ditlevsen}
P. Ditlevsen, and I. A. Mogensen, Phys. Rev. E, {\bf 53}, 4785, (1996).
\bibitem{PD}
P. Ditlevsen, private communication.
\end{thebibliography}
\end{document}